\documentclass{article}
\usepackage{graphics}
 \usepackage{graphicx}

 \usepackage{epsfig}

\usepackage{amssymb}
\usepackage{amsmath}
 \usepackage{amsthm}

\begin{document}

\title{Photon-phonon-assisted thermoelectric effects in the molecular devices}

\author{M.˜ Bagheri Tagani,
        H.˜ Rahimpour Soleimani,\\
        \small{Department of physics, University of Guilan, P.O.Box 41335-1914, Rasht, Iran}}

\maketitle

\begin{abstract}
We consider a single level quantum dot interacting with a phonon
mode and weakly coupled to metallic leads which are subjected to a
time-dependent gate voltage. Electrical conductance, thermopower,
and figure of merit are investigated in detail using a
Tien-Gordon-type rate equation. In the presence of the microwave
field, the electrical conductance exhibits extra peaks whose
height is controlled by the magnitude of the microwave field.
Furthermore, the oscillation of the thermopower increases in the
presence of the time-dependent gate voltage or the
electron-phonon interaction. Influence of the electron-phonon
interaction, microwave field, temperature, and Coulomb
interaction on the figure of merit is also studied.
\end{abstract}

\section{Introduction}
\label{Introduction} The study of the thermoelectrical properties
of nanoscale devices has attracted a lot of attention in recent
years theoretically and
experimentally~\cite{ref1,ref2,ref3,ref4,ref5,ref6,ref7,ref8,ref9,ref10,ref11,ref12}.
The violation of the Wiedemann– Franz law~\cite{ref13,ref14}, that
the ratio of the electrical conductance to the thermal
conductance is proportional to the temperature, in nanostructures
has caused that the nanostructures have been considered as good
candidates for fabricating thermoelectric devices. The
thermoelectric efficiency is measured by a dimensionless quantity
as figure of merit, $ZT=S^2G_eT/(\kappa_e+\kappa_{ph})$ where $S$
stands for the thermopower and $G_e$ denotes the electrical
conductance. $\kappa_e$, and $\kappa_{ph}$ are the electrical and
phononic thermal conductances, respectively. The Coulomb blockade
effect~\cite{ref13,ref15}, discreteness of the energy
levels~\cite{ref16}, interference effects~\cite{ref10} and so on
are the main reasons for high figure of merit in nanoscale
devices. The quantum dot (QD) as a promising thermoelectric
material has been widely studied during recent
years~\cite{ref3,ref4,ref11,ref12,ref13,ref15,ref16,ref17,ref18,ref19,ref20,ref21,ref22,ref23,ref24,ref25}.
\par The influence of the time-dependent gate voltage on the
thermoelectric properties and on the heat transport through
QD-based devices is a very interesting issue which has been
considered very recently. It was predicted that the heat can be
conducted by phonons in the low temperature~\cite{ref26}. In
addition, it was reported that the thermopower can be enhanced in
the presence of a step-like pulse bias~\cite{ref27,ref28}. Pei and
co-workers~\cite{ref29} studied the transient heat generation in a
QD by applying a step-like pulse bias. They found that the
time-dependent bias results in the periodic oscillation of the
heat generation. Chi and co-workers~\cite{ref30} studied the
thermoelectric properties of a QD in the presence of a microwave
field. They reported that the applying microwave leads to the
appearance of the extra peaks in the electrical conductance
spectrum.
\par The coupling between the electronic degree of freedom and
vibrational degree of freedom or the electron-phonon interaction
(EPI) is an important issue in the transport through nanoscale
devices resulting in the novel and interesting phenomena. The
influence of the EPI on the transport properties of the molecular
transistors and the carbon nanotube quantum dots has been
extensively studied using both the Keldysh nonequilibrium Green
function formalism and the rate equation
approach~\cite{ref31,ref32,ref33,ref34,ref35,ref36,ref37,ref38,ref39,ref40,ref41,ref42}.
In addition, the thermoelectric properties of the molecular
transistors has been analyzed in recent
years~\cite{ref12,ref43,ref44,ref45,ref46}. However, although the
transport properties of such devices has been studied in the
presence of the EPI and the time-dependent
voltages~\cite{ref47,ref48}, the influence of the time-dependent
gate voltage on the thermoelectric properties of the molecular
devices has not been addressed so far. In this article, we
consider a molecule attached to the metallic leads which are
subjected to a microwave field in the non-adiabatic regime. We
follow the method introduced by Dong and co-workers~\cite{ref48}
to obtain the formal expressions of the thermopower and electrical
conductance. The analytical relations for the thermopower and
electrical conductance are obtained using a Tien-Gordon-type rate
equation. The influence of the EPI, microwave field, temperature
and Coulomb repulsion on the figure of merit is also examined. The
model and formalism are presented in the next section. Section 3
is devoted to the numerical results, and in the end, some
sentences are given as a summary.

 \section{Model and formalism}
 \label{Model}
We consider a single level QD interacting with a dispersionless
optical phonon mode localized in the QD. The QD is weakly coupled
to the normal metal electrodes which are subjected to a
time-dependent gate voltage. The Hamiltonian describing the whole
system is given as
\begin{align}\label{Eq.1}
  H&=\sum_{\alpha k\sigma}\varepsilon_{\alpha
  k\sigma}(t)c^{\dag}_{\alpha k\sigma}c_{\alpha
  k\sigma}+\sum_{\sigma}\varepsilon_{\sigma}n_{\sigma}+Un_{\uparrow}n_{\downarrow}+\omega
  a^{\dag}a+\\\nonumber
  &\lambda\sum_{\sigma}[a^{\dag}+a]n_{\sigma}+\sum_{\alpha
  k\sigma}[V_{\alpha k\sigma}c^{\dag}_{\alpha
  k\sigma}d_{\sigma}+H.c.],
\end{align}
where $c_{\alpha k\sigma}(c^{\dag}_{\alpha k\sigma})$ destroys
(creates) an electron with wave vector $k$, spin $\sigma$ in lead
$\alpha$. $\varepsilon_{\alpha k\sigma}(t)=\varepsilon^{0}_{\alpha
k\sigma}+u_{\alpha}cos\Omega t$ denotes the energy levels of the
leads composed of the rigid shift of the Fermi energies of the
leads, $\varepsilon^{0}_{\alpha k\sigma}$, and the time-dependent
energies, $u_{L(R)}=\pm V_{ac}$ induced by a microwave field with
the frequency $\Omega$ and the magnitude $V_{ac}$. $d_{\sigma}
(d^{\dag}_{\sigma})$ is the annihilation (creation) operator in
the QD and $n_{\sigma}=d^{\dag}_{\sigma}d_{\sigma}$ is the
occupation operator. $U$ denotes the on-site Coulomb repulsion
whereas, $\omega$ is the phonon energy. $\lambda$ and $V_{\alpha
k\sigma}$ describe the electron-phonon coupling strength and
tunneling between the QD and the leads, respectively.
\par The EPI can be eliminated by using a polaronic
transformation~\cite{ref49}, $\tilde{H}=e^{S}He^{-S}$ where
$S=exp((a^{\dag}-a)\sum_{\sigma}n_{\sigma})$. The transformation
results in decoupling the electronic and phononic subsystems and
the renormalization of the QD energy levels and the Coulomb
repulsion. The transformed Hamiltonian is given as
\begin{align}\label{Eq.2}
  \tilde{H}&=\sum_{\alpha k\sigma}\varepsilon_{\alpha k\sigma}
  c^{\dag}_{\alpha k\sigma}c_{\alpha
  k\sigma}+\sum_{\sigma}\tilde{\varepsilon}_{\sigma}n_{\sigma}+\tilde{U}n_{\uparrow}n_{\downarrow}\\\nonumber
  &\omega a^{\dag}a+\sum_{\alpha k\sigma}[V_{\alpha k\sigma}Xc^{\dag}_{\alpha k\sigma}d_{\sigma}+H.c.]
\end{align}
where
$\tilde{\varepsilon}_{\sigma}=\varepsilon_{\sigma}-\lambda^2/\omega$,
$\tilde{U}=U-2\lambda^2/\omega$, and
$X=exp(-\lambda/\omega(a^{\dag}-a))$. In the following, we use
the wide band approximation so that the tunneling rate,
$\Gamma^{\alpha}_{\sigma}=\sum_{k}|V_{\alpha k\sigma}|^2$, is
energy-independent. Because the QD-lead coupling is assumed to be
so weak, $\Gamma^{\alpha}_{\sigma}<<T$, where $T$ denotes the
temperature, and on the other hand, the frequency of the
microwave field is so stronger than the tunneling rate, a
Tien-Gordon-type rate equation~\cite{ref48,ref50} can be used to
describe the behavior of the system. With respect to
Eq.~\eqref{Eq.2}, the isolated QD can be in the empty state,
one-electron state, or doubly occupied state with $n$ phonons.
These states are shown by $P_{N}^{n}$ where $N$ and $n$ stand for
the number of the electrons and phonons, respectively, and obey
the following rate equations:
\begin{subequations}\label{Eq.3}
\begin{align}
\frac{P_0^n}{dt}&=\sum_{\alpha m\sigma}[\Gamma_{\alpha \sigma
0}^{mn}P_{\sigma}^{m}-\Gamma_{\alpha 0\sigma}^{nm}P_{0}]\\
\frac{P_{\sigma}^{n}}{dt}&=\sum_{\alpha m}[\Gamma_{\alpha
0\sigma}^{mn}P_{0}^{m}+\Gamma_{\alpha
2\sigma}^{mn}P_{2}^{m}-(\Gamma_{\alpha\sigma
0}^{nm}+\Gamma_{\alpha\sigma 2}^{nm})P_{\sigma}^{n}]\\
\frac{P_{2}^{n}}{dt}&=\sum_{\alpha m\sigma}[\Gamma_{\alpha\sigma
2}^{mn}P_{\sigma}^{m}-\Gamma_{\alpha 2\sigma}^{nm}P_{2}^{n}]
\end{align}
\end{subequations}
where $\Gamma_{\alpha NM}^{nm}$ describes the transition from
N-electron state to M-electron state, while the number of phonons
is changed from $n$ to $m$. The transition rates are defined
as~\cite{ref48}
\begin{align}\label{Eq.4}
\Gamma_{\alpha
NM}^{nm}&=\Gamma^{\alpha}e^{-g^2}g^{2|m-n|}\frac{p!}{q!}(L_{p}^{|m-n|}
(g^2))^2\sum_{j=-\infty}^{\infty}J_j(u_\alpha/\Omega)^2\\\nonumber
&[f_{\alpha}(\tilde{\varepsilon}_M-\tilde{\varepsilon}_N+(m-n)\omega-j\Omega)\Theta(M-N)+\\\nonumber
&(1-f_{\alpha}(\tilde{\varepsilon}_N-\tilde{\varepsilon}_M+(n-m)\omega-j\Omega))\Theta(N-M)]
\end{align}
where $g=\lambda/\omega$, $p=min[m,n]$, $q=max[n,m]$, and
$\Theta(x)$ is the Heaviside step function. $N$ and $M$ denote
the number of the electrons in the QD expressed as $N=0$, empty
state, $N=1$, $\tilde{\varepsilon_N}=\varepsilon_\sigma$ (single
electron state), and $N=2$,
$\tilde{\varepsilon_N}=\tilde\varepsilon_\uparrow+\tilde\varepsilon_\downarrow+\tilde{U}$
(two electron state). $L_{p}^{q}(x)$ is the generalized Laguerre
polynomial and $J_j(x)$ is the Bessel function of order $j$.
$f_{\alpha}(x)=(1+exp((x-\mu_\alpha)/kT))^{-1}$ is the Fermi
distribution function of lead $\alpha$  and $\mu_\alpha$ stands
for the chemical potential of the lead. Note that the leads are
assumed to be the normal metal, therefore, the tunneling rate,
$\Gamma^\alpha$, is spin-independent.
\par Solving Eqs.~\eqref{Eq.3} in the steady state
($dP_N^n/dt=0$), the charge and energy currents are expressed as
\begin{subequations}\label{Eq.5}
\begin{align}
I^{\alpha}&=-\frac{e}{\hbar}\sum_{\sigma nm}[\Gamma^{nm}_{\alpha
0\sigma}P_{0}^{n}+(\Gamma^{nm}_{\alpha\sigma
2}-\Gamma^{nm}_{\alpha\sigma
0})P_{\sigma}^{n}-\Gamma^{nm}_{\alpha2\sigma}P_2]\\
Q^{\alpha}&=\frac{1}{\hbar}\sum_{\sigma nm}[\Gamma^{Qnm}_{\alpha
0\sigma}P_{0}^{n}+(\Gamma^{Qnm}_{\alpha\sigma
2}-\Gamma^{Qnm}_{\alpha\sigma
0})P_{\sigma}^{n}-\Gamma^{Qnm}_{\alpha2\sigma}P_2]
\end{align}
\end{subequations}
where $\Gamma^{Qnm}_{\alpha NM}$ describes the energy transported
in the tunneling process and expressed as
\begin{align}\label{Eq.6}
\Gamma_{\alpha
NM}^{Qnm}&=\Gamma^{\alpha}e^{-g^2}g^{2|m-n|}\frac{p!}{q!}(L_{p}^{|m-n|}
(g^2))^2\sum_{j=-\infty}^{\infty}J_j(u_\alpha/\Omega)^2\\\nonumber
&[(\tilde{\varepsilon}_M-\tilde{\varepsilon}_N+(m-n)\omega-j\Omega)f_{\alpha}(\tilde{\varepsilon}_M-\tilde{\varepsilon}_N+(m-n)\omega-j\Omega)\Theta(M-N)+\\\nonumber
&(\tilde{\varepsilon}_N-\tilde{\varepsilon}_M+(n-m)\omega-j\Omega)(1-f_{\alpha}(\tilde{\varepsilon}_N-\tilde{\varepsilon}_M+(n-m)\omega-j\Omega))\Theta(N-M)]
\end{align}
\par In the linear response regime, the charge and heat currents
are expanded as~\cite{ref49}
\begin{subequations}\label{Eq.7}
\begin{align}
I^{\alpha}&=G_e\Delta V+G_T\Delta T \\
Q^{\alpha}&=M\Delta V+K\Delta T
\end{align}
\end{subequations}
where $G_e$ and $G_T$ are the electrical conductance and thermal
coefficient, respectively. The thermopower is computed as
$S=-\Delta V/\Delta T$ in the limit of zero current. In order to
compute the above coefficients, we assume $T_L=T+\Delta T$,
$\mu_L=Ef+e\Delta V$, $\mu_R=Ef$, and $T_R=T$ where $Ef$ denotes
the Fermi energy of the leads. By expanding the Fermi function of
the left lead according to
$f_L(\varepsilon)=f(\varepsilon)-e\Delta
Vf'(\varepsilon)-\frac{\Delta
T}{T}(\varepsilon-Ef)f'(\varepsilon)$ where
$f'(\varepsilon)=\partial{f(\varepsilon)}/\partial(\varepsilon)$
and setting $I^L=1/2(I^L-I^R)$ and $Q^L=1/2(Q^L-Q^R)$, one can
easily obtain the thermoelectric coefficients. For simulation
purpose, we set $\omega=1$ as the energy unit. The reported value
of the phonon energy is varied from $10\mu{eV}$ to $10meV$ in the
literatures. In addition, we set $\kappa_{ph}=3\kappa_0$  where
$\kappa_0=\pi^2k^2/3hT$  is the quantum of thermal
conductance~\cite{ref51} and, assume that the energy level of the
QD is degenerate.

%\begin{figure}[htb]
%\begin{center}
%\includegraphics[height=100mm,width=100mm,angle=0]{fig1}
%\caption{(a) Electrical conductance as a function of electron
%temperature for $\lambda=0$ (solid line), $\lambda=1,T_p=T_e$
%(dashed line), and $T_p=5T_e$ (dash-dotted line). (b)-(d) show DOS
%in the parts1-3, respectively.
%$-\frac{\partial{f(\varepsilon)}}{\partial{\varepsilon}}$ is
%plotted in gray. $\Gamma^{i}_{\alpha\sigma}=0.2\omega$, and
%$\Delta=2\omega$ is the level spacing.}\label{fig:1}       % Give a unique label
%\end{center}
%\end{figure}

\section{Numerical results and discussions }
\label{Numerical results}

Figs. 1a and 1b describe the electrical conductance as a function
of the energy level of the QD and temperature, respectively. As
it is expected, the main peaks of the electrical conductance are
located in the resonance energies, i.e., $\tilde{\varepsilon}=0$,
and $\tilde{\varepsilon}=-\tilde{U}$. First, we analyze the
behavior of $G_e$ in the absence of the EPI. By applying
microwave field the electrical conductance shows extra peaks at
the energies $\tilde{\varepsilon}=\pm n\Omega$, and
$\tilde{\varepsilon}=-U\pm n\Omega$ where $n$ is an integer.
These side peaks are induced by the photon-assisted tunneling
channels. Such result has been recently reported in
Ref.~\cite{ref30}. By solving Eqs.~(4,7), we have obtained the
following analytical relation for the electrical conductance
\begin{align}\label{conductance}
  G_e&=\frac{e^2}{\hbar}\frac{\Gamma^L}{2kTZ}[(1+e^{-\varepsilon/kT})\sum_{n=-\infty}^{\infty}\frac{J_n(u_L/\Omega)^2}{1+cosh((Ef+n\Omega-\varepsilon)/kT)}+\\\nonumber
  &(e^{-\varepsilon/kT}+e^{-(2\varepsilon+U)/kT})\sum_{n=-\infty}^{\infty}\frac{J_n(u_L/\Omega)^2}{1+cosh((Ef+n\Omega-\varepsilon-U)/kT)}]\nonumber
\end{align}
where $Z=1+2exp(-\varepsilon/kT)+exp(-(2\varepsilon+U)/kT)$ is
the partition function and $J_n(x)$ is the Bessel function of
order $n$. It is straightforward to show that the electrical
conductance obeys the following equation for the case $V_{ac}=0$
\begin{equation}\label{conductance1}
  G_e=\frac{e^2}{\hbar}\frac{\Gamma^L}{kTZ}[\frac{1}{1+e^{\varepsilon/kT}}+\frac{e^{U/kT}+e^{-\varepsilon/kT}}{(1+e^{(\varepsilon+U)/kT})^2}]
\end{equation}
that it is in agreement with the results presented in
Ref.~\cite{ref43}. According to Eq.~\eqref{conductance1}, the
height of the main peaks of the electrical conductance is equal to
$\frac{e^2}{\hbar}\frac{\Gamma^L}{6kT}$ if $u_L=0$. In the
presence of the time-dependent voltage, the height of the main
peaks decreases proportional to $J_0(u_L/\Omega)^2$. Therefore,
the ratio of the main peak's height in  the case $u_L\neq 0$ to
$u_L=0$ can contain the important information about the influence
of the time-dependent gate voltage on the energy levels of the
electrodes. Moreover, the appearance of the extra peaks in the
electrical conductance is explained by Eq.~\eqref{conductance}.
The height of these side peaks is controlled by
$J_{n}(u_L/\Omega)^2$ so that  their height increases with
increase of $u_L$. The dependence of the electrical conductance
on the temperature is strongly dependent on the energy level of
the QD, so that the electrical conductance is significantly
reduced in the resonance energies with increase of temperature.
The behavior of the electrical conductance in the electron-hole
symmetry point, $\tilde{\varepsilon}=-\tilde{U}/2$, is very
interesting. In low temperatures, $kT<<U$, and in the absence of
the microwave fields, the electrical conductance is zero because
the electrons and holes participate in the current with the same
weight. The magnitude of the electrical conductance significantly
increases in the presence of the time-dependent voltage. By
analyzing Eq.~\eqref{conductance} in the low temperature limit,
it is found that the electrical conductance is equal to
$\frac{e^2}{\hbar}\frac{\Gamma^L}{8kT}[J_{-1}(u_L/\Omega)^2+J_{1}(u_L/\Omega)^2]$
so the magnitude of $G_e$ is increased with increase of $u_L$.
The influence of the photons is suppressed in the higher
temperatures so that the electrical conductance becomes
independent of the photon-assisted channels. In the high
temperature, the various electronic states participate in the
charge transport through the QD thus, the extra channels induced
by photons do not have significant role in the charge transport.
Now, we study the effect of the EPI on the electrical
conductance. As one expects, the position of the main peaks is
now shifted because of the polaronic shift. It is important to
note that the distance between the peaks is controlled by the
Coulomb interaction considered to be constant in the article,
$\tilde{U}=2$. The height of the peaks is exponentially reduced
in the presence of the EPI. A careful analysis reveals that the
reduction is proportional to the EPI strength as $e^{-g^2}$. Such
decrease was also observed in the Kondo regime~\cite{ref52} owing
to the existence of factor $e^{-g^2}$  in the density of the
states of the QD. Here, the reduction is caused due to the
exponential suppression of the tunneling between the QD and the
leads in the presence of the EPI. Results also show that the EPI
results in the reduction of the electrical conductance, more
specifically, in high temperatures. However, the
photon-phonon-assisted electrical conductance is more than the
elastic case in low temperatures. The decrease of the
phonon-assisted electrical conductance comes from the fact that
the electron configurations with few phonons are now partially
populated but their participations in the conductance are
suppressed by $|L_{m>0}^{n}(g^2)|^2$.
\par Fig. 2 describes the thermopower as a function of the energy
level of the QD. The sawtooth oscillation of the thermopower is
risen due to the change of the electron number in the QD. By
solving Eqs.~(4,7), the following analytical relation for the
thermopower is obtained in the case $\lambda=0$
\begin{align}\label{thermopower}
  S&=\frac{-e\Gamma^L}{2\hbar kT^2G_eZ}[(1+e^{-\varepsilon/kT})\sum_{n=-\infty}^{\infty}\frac{J_n(u_L/\Omega)^2(\varepsilon-n\Omega)}{1+cosh((Ef+n\Omega-\varepsilon)/kT)}+\\\nonumber
  &
  (e^{-\varepsilon/kT}+e^{-(2\varepsilon+U)/kT})\sum_{n=-\infty}^{\infty}\frac{J_n(u_L/\Omega)^2(\varepsilon+U-n\Omega)}{1+cosh((Ef+n\Omega-\varepsilon-U)/kT)}]\nonumber
\end{align}
When $u_L=0$, the thermopower has three zeros located in the
resonance energies and electron-hole symmetry point. In resonance
energies the temperature gradient cannot produce the net current
and, as a result, the thermopower becomes zero. In the
electron-hole symmetry point, both electrons and holes carry the
charge and energy with the same weight. Although they carry the
energy in the same direction, they carry the charge in the
opposite directions so that no net current is produced. The sign
of the thermopower shows the kind of charge participated in the
transport so that $S>0$ is for holes, while $S<0$ is for
electrons. In the energies more than the resonance energies
electrons injected from the hotter lead (the left lead) create
the current, whereas for the right hand side of the electron-hole
symmetry point, holes create the current. Such behavior was
previously reported for single QD and double quantum dot
systems~\cite{ref16,ref23,ref24}. In the presence of the microwave
fields and EPI, the oscillations of the thermopower increase
because of the phonon-photon-assisted tunneling channels.
However, the magnitude of the thermopower decreases proportional
to $J_n(u_L/\Omega)^2$. Moreover, oscillations become faster with
increase of $u_L$, because the energy levels of the leads
oscillate faster. The slope of the thermopower is roughly
estimated like $1/T$ which is in agreement with the results
presented in Ref.~\cite{ref43} for single molecule devices in the
sequential tunneling regime. In the energies near the chemical
potential of the leads, the sign of the thermopower changes more
because the phonons and photons have more meaningful role in the
charge transport from the leads to the QD.

\par The dependence of the figure of merit on the energy level of
the QD and temperature is shown in figs. 3a and 3b, respectively.
The figure of merit is zero  in the resonance energies and
electron-hole symmetry points because of $S=0$. The EPI and
microwave fields result in the reduction of the magnitude of the
figure of merit because they modulate the tunneling process
between the QD and the electrodes so that the thermopower and
electrical conductance decrease.   However, the number of peaks of
$ZT$ significantly increases in the presence of the microwave
fields or the EPI owing to the photon- or phonon-assisted
channels. Indeed, the time-dependent gate voltage results in the
increase of the thermoelectric efficiency of the device in
energies in which the device cannot work as a thermoelectric
device without the microwave fields. The same result was reported
by  Chi and co-workers~\cite{ref30}. In the low temperatures, the
figure of merit has a maximum (fig. 3b) because the Coulomb
interaction controls the transport process. With increase of
temperature resulting in the reduction of the Coulomb interaction
effect, the figure of merit is reduced. The position of the peak
is slightly shifted in the presence of the EPI or the microwave
fields.

\par Fig. 4a shows the dependence of $ZT$ on the strength of the
EPI and the magnitude of the microwave field. Results show that
applying the microwave can partly compensate the negative effect
of the EPI on the figure of merit. It comes from the
photon-assisted channels resulting in the increase of the
thermopower and electrical conductance even in the presence of
the EPI. Therefore, the time-dependent gate voltage can be used
to improve the thermoelectric efficiency of the molecular devices
in the presence of the EPI. The influence of the Coulomb
interaction on the figure of merit is analyzed in fig. 4b. In the
range of $\tilde{U}\cong -2\tilde{\varepsilon}$,
($\tilde{\varepsilon}=-0.5$), $ZT$ is zero because of the
electron-hole symmetry point. Indeed, $ZT$ approaches zero when
the electrons and holes participate in the charge and energy
transport. In the both weak and strong $U$s, $ZT$ takes the
finite values. In these cases, electrons carry the charge and
energy through the system because the QD is in the two-electron
state or one-electron state, respectively. It is interesting to
note that $ZT$ takes the reasonable values in high $\lambda$s
only in strong $U$s. It comes from the fact that the empty state
ore two-electron state will be populated in strong $\lambda$ and
nearly weak $U$, so that holes can also participate in the
transport resulting the decrease of the thermopower. Therefore,
$ZT$ becomes very small in weak $U$ and strong $\lambda$.

%\begin{figure}[htb]
%\begin{center}
%\includegraphics[height=80mm,width=80mm,angle=0]{fig6}
%\caption{ Dependence of Lorentz ratio on the electron
%temperature.}\label{fig:6}       % Give a unique label
%\end{center}
%\end{figure}
\section{Summary}
\label{Summary} In this article, we have investigated the
thermoelectric properties of the molecular devices in the
presence of the time-dependent voltage by means of a
Tien-Gordon-type rate equation formalism. Analytical expressions
of the thermopower and electrical conductance are obtained.
Results show that the microwave field results in the appearance
of the extra peaks in the conductance spectrum whose height is
controlled by the magnitude of the microwave. Electron-phonon
interaction results in the reduction of the figure of merit.
However, the microwave field can slightly compensate the negative
effect of the EPI. We also find that the figure of merit takes
reasonable values in the strong Coulomb interaction if the EPI is
very strong.

\bibliographystyle{model1a-num-names}
\bibliography{<your-bib-database>}

\newpage
\textbf{Figure captions}

\par Figure 1: Electrical conductance (a) as a function of the
energy level of the QD, (b) as a function of temperature.
Parameters are $\Omega=0.5$, $\Gamma_0=0.01\omega$, and
$\tilde{U}=2\omega$. $kT=0.1$in fig. 1a and $\varepsilon=-\omega$
in fig. 1b.

\par Figure 2: Thermopower versus the energy level of
the QD. Parameters are the same as fig. 1 except
$\tilde{\varepsilon}=-0.5\omega$.

\par Figure 3: Dependence of $ZT$ on (a) energy level of the QD,
(b) temperature. Parameters are the same as fig. 1.
$\tilde{\varepsilon}=-0.5\omega$ in fig. 3b.

\par Figure 4: Color map of $ZT$. Parameters are the same as fig.
3. $V_{ac}=\omega$ in fig. 4b.

\newpage

\begin{figure}[htb]
\begin{center}
\includegraphics[height=150mm,width=150mm,angle=0]{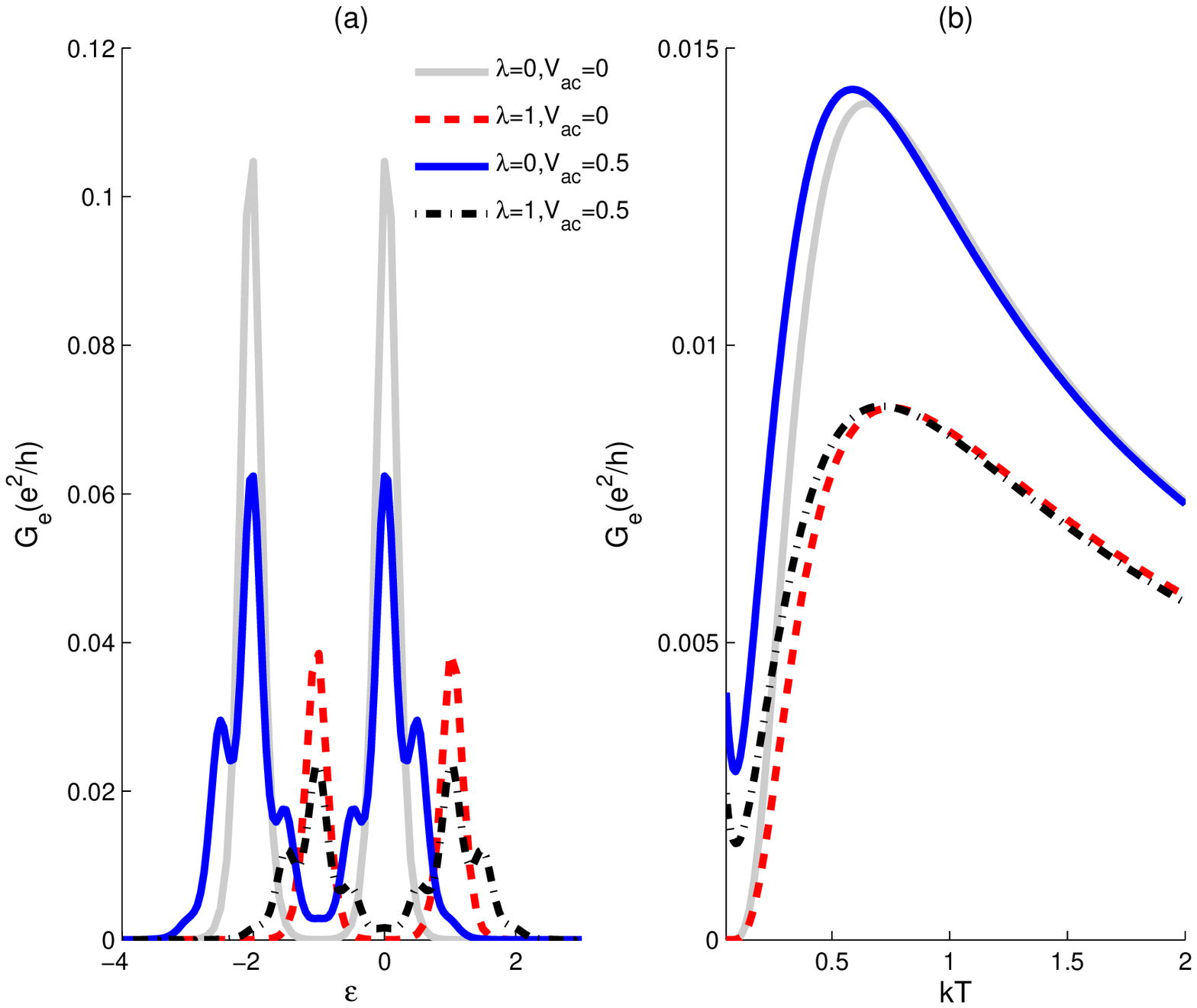}\nonumber
$Figure. 1$
\end{center}
\end{figure}

\begin{figure}[htb]
\begin{center}
\includegraphics[height=150mm,width=150mm,angle=0]{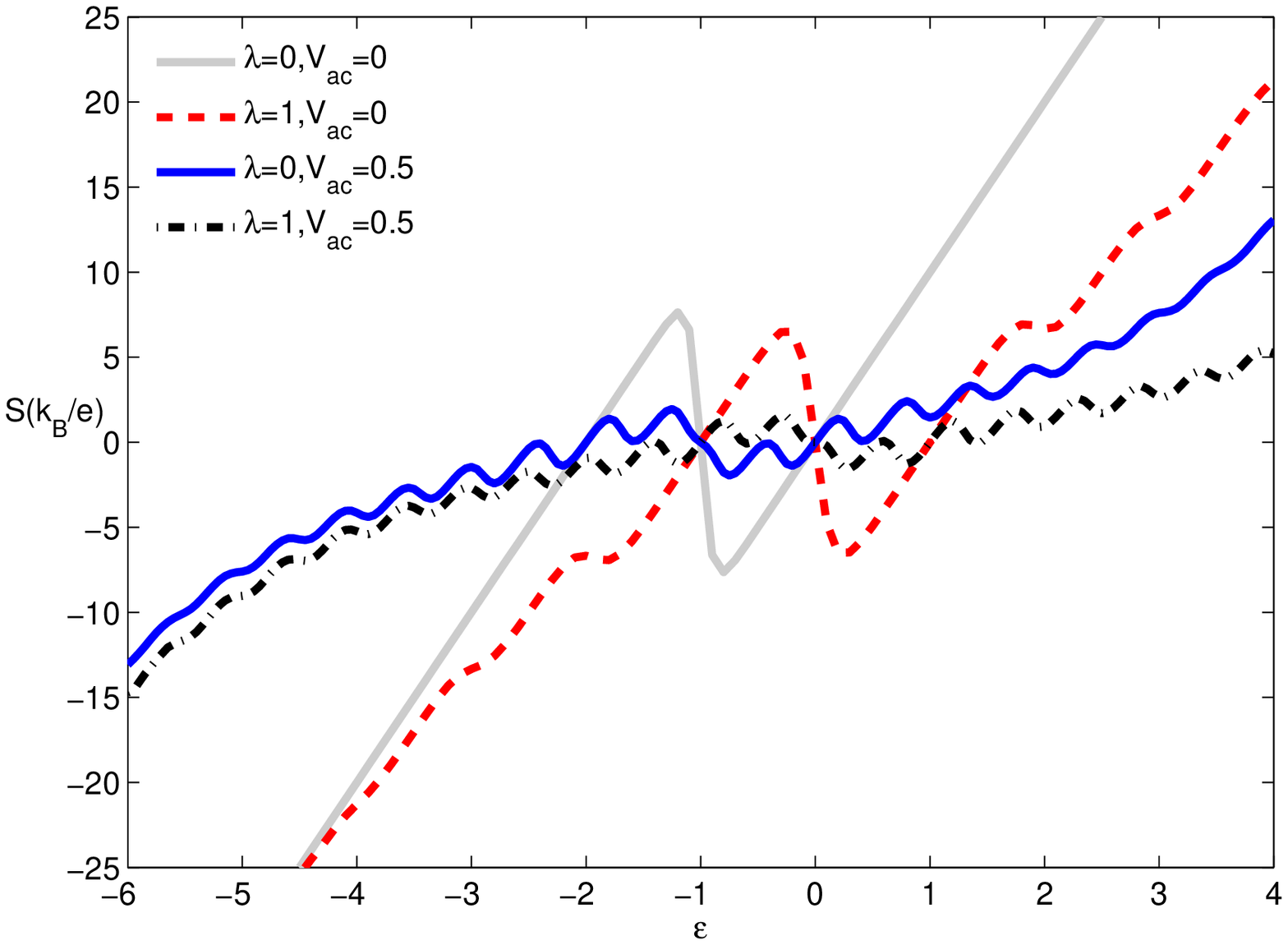}\nonumber
$Figure. 2$\end{center}
\end{figure}

\begin{figure}[htb]
\begin{center}
\includegraphics[height=150mm,width=150mm,angle=0]{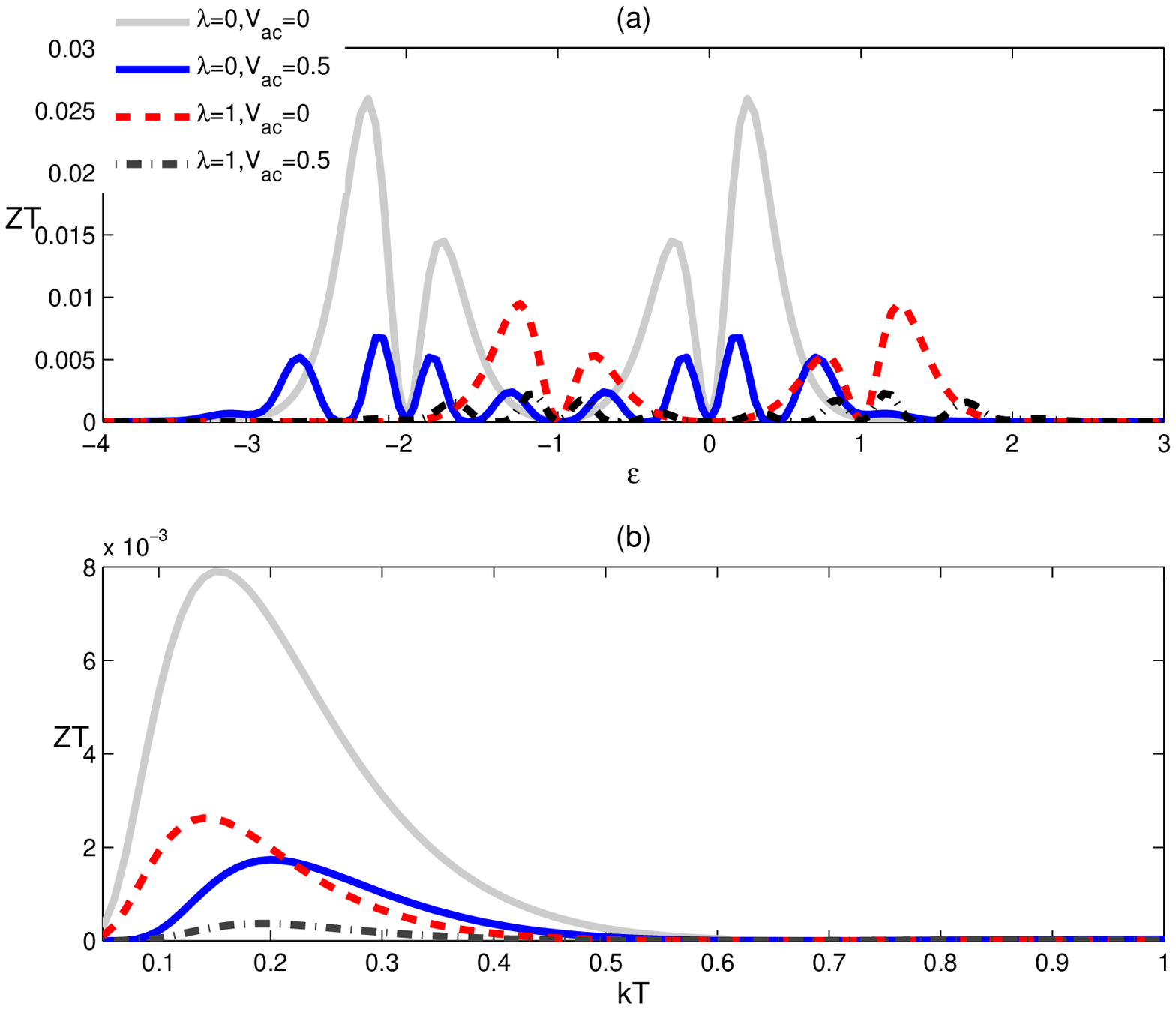}\nonumber
$Figure. 3$
\end{center}
\end{figure}

\begin{figure}[htb]
\begin{center}
\includegraphics[height=150mm,width=150mm,angle=0]{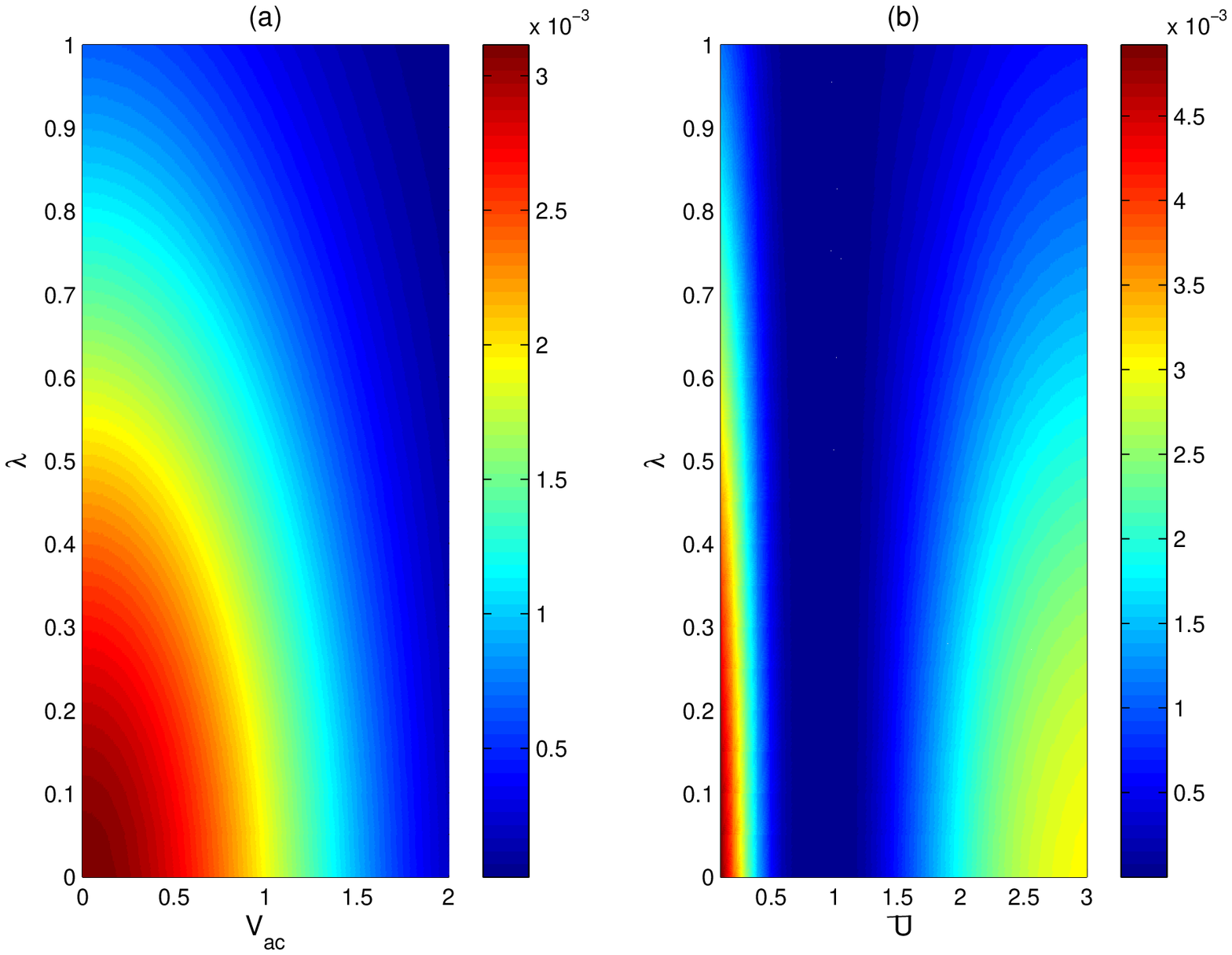}\nonumber
$Figure. 4$
\end{center}
\end{figure}

\end{document}